\documentclass[%
 reprint,
 amsmath,amssymb,
 aps,
prstab,
physics,
longbibliography
]{revtex4-1}

\usepackage{graphicx}
\usepackage{dcolumn}
\usepackage{bm}

\begin{document}

\preprint{APS/123-QED}

\title{Compact Source of Positron Beams with Small Thermal Emittance}

\author{Rafi Hessami}
\email{rafimah@stanford.edu}
 \affiliation{Applied Physics Department, Stanford University, Stanford, USA}
 \affiliation{SLAC National Accelerator Laboratory, Menlo Park, USA}
 
\author{Spencer Gessner}
\email{sgess@slac.stanford.edu}
\affiliation{SLAC National Accelerator Laboratory, Menlo Park, USA}

\date{\today}

\begin{abstract}
We investigate electrostatic traps as a novel source of positron beams for accelerator physics applications. Penning-Malmberg (PM) traps are commonly employed in low-energy antimatter experiments. Positrons contained in the trap are cooled to room temperature or below. We calculate the thermal emittance of the positrons in the trap and show that it is comparable to or better than the performance of state-of-the-art photocathode guns. We propose a compact positron source comprised of a PM trap, electrostatic compressor, and rf accelerator that can be built and operated at a fraction of the cost and size of traditional target-based positron sources, albeit at reduced repetition rate and with intrinsic angular momentum. We model the acceleration of a positron bunch up to an energy of 17.6 MeV with a final thermal emittance of 0.60 $\mu$m rad and bunch length of 190 $\mu$m. This system may be useful for accelerator physics studies, such as investigations of flat-beam sources for linear colliders and positron plasma wakefield acceleration. 

\end{abstract}

\pacs{Valid PACS appear here}
\maketitle

\section{\label{intro} Introduction}
Positron beams are traditionally produced by sending high-energy electron beams into a high-Z target, capturing positrons from the resulting electromagnetic shower, and cooling the positrons in a damping ring before reacceleration~\cite{Chaikovska2022}. This process requires significant experimental infrastructure and hardware. As a result, there are relatively few laboratories producing positron beams for accelerator physics experiments~\cite{Yakimenko2019}. Research into advanced positron sources has been recognized as an area-of-need for future accelerator R\&D~\cite{snowmassPosi}.

One research area impacted by the lack of positron beam sources is Plasma Wakefield Acceleration (PWFA). PWFA is a promising technique for accelerating charged particles at high gradients. Preserving the quality of positron beams while accelerating them in plasma is an unsolved challenge~\cite{Hogan2003,Blue2003,Muggli2008,Corde2015,Gessner2016,Doche2017}. The question of how best to accelerate a positron beam in plasma can only be resolved by committing significant experimental and computational resources to the task. New types of positron sources will expand access to positron beams which can be used for these experiments.

We propose a novel, compact, electrostatic positron source for accelerator physics research. Previous research has explored electron beams from ulta-cold plasmas (UCP)~\cite{Claessens2005,Taban2008} and magneto-optical traps (MOT)~\cite{Xia2014,Franssen2019}. Our concept is the first to examine this possibility for positron beams. The positron source is based on the electrostatic Penning-Malmberg (PM) trap, commonly employed in low-energy antimatter experiments~\cite{RevModPhys.87.247}. These traps have the advantage of providing cold, low-emittance beams, with the caveat that the repetition rate of these devices is low and the beams have intrinsic angular momentum. The trap is combined with a short linac to compress and accelerate the beam such that the final energy and bunch length is suitable for injection in a plasma wake. While positron PWFA experiments are the motivation for this concept, the compact positron source would be of great interest to any facility that desires positron beams for physics studies.

\section{Overview of the Electrostatic Positron Beam Source}

In this section, we provide a description of the electrostatic beam source and explain how properties of the electrostatic trap impact beam parameters like bunch length and emittance. The review of electrostatic traps by Danielson, et. al. provides a detailed overview of these systems~\cite{RevModPhys.87.247}.

\subsection{Positron Sources}

Positrons for electrostatic traps are typically produced by $\beta$-decay emitters such as $^{22}$Na. The emitters are sold as small encapsulated sources that can be attached to a vacuum beamline. The primary limitation of the encapsulated source is that they contain a limited amount of radioactive material for safe handling and produce at most $10^9$ positrons per second~\cite{KrauseRehberg2004}.

An alternative method for generating positrons for the compact source employs a small, 9 MeV electron accelerator and impacts the beam on a high-Z target~\cite{Charlton2021}. This creates low-energy positrons from an electromagnetic shower, but the initial beam energy is low enough as to not activate the target material which reduces shielding requirements. This approach is being pursued by the GBAR experiment at CERN with a goal $10^{10}$ positrons per second from the target~\cite{Charlton2021}.

For both the encapsulated radioactive source and the compact accelerator-based source, the positrons have a large kinetic energy relative to the depth of the electrostatic trap and a large energy spread. In order to trap the positrons, the beam must first be sent through a moderator which slows the positrons. A commonly employed moderator is solid neon with an efficiency of $10^{-2}$~\cite{doi:10.1063/1.97441}. Therefore, the flux of slow positrons into the trap is about $10^7$ positrons per second for an encapsulated radioactive source and $10^8$ positrons per second for the accelerator-based source.

\subsection{The Electrostatic Trap}

The positrons enter an electrostatic trap consisting of a series of ring electrodes surrounded by a solenoid magnet. The ring electrodes create the axial potential well that traps the positrons longitudinally, while the solenoid provides radial confinement. The depth of the well needs to be greater than the space charge potential of the positrons in the trap, given by
\begin{equation}\label{eq:depth}
   \Delta \phi = \frac{e n r_p^2}{4\varepsilon_0}\left[1+2\ln{\left(\frac{r_w}{r_p}\right)}\right],
\end{equation}
for positron density $n$, plasma radius $r_p$, and trap radius $r_w$~\cite{RevModPhys.87.247}.

The properties of the beam inside the electrostatic trap are defined by the trap's parameters. In particular, the radial extent of the positrons in the trap, and therefore the density of the positrons in the trap are defined by the magnetic field and the rotation rate of the positron plasma. The rotation rate is a free parameter which can be imposed upon the positron plasma through a ``rotating wall" electrode~\cite{PhysRevLett.85.1883}. In this scenario, the positron plasma is a uniform cylinder of charge extending to radius $r_p$ with the density given by
\begin{equation} \label{dens_omega}
    n = \frac{2\varepsilon_0 B \omega_r}{e},
\end{equation}
where $B$ is the solenoid field and $\omega_r$ is the rotation rate of the positron plasma.

Our calculations and simulations assume trap parameters that have previously been achieved in experiments. The trap parameters and beam parameters used in the simulation are shown in Table~\ref{tab:parameters}. We note that the parameters we chose for our simulation are conservative. For example, we assume a solenoid field of 1 T whereas the GBAR experiment employs a 5 T magnet~\cite{GBARproposal}, and a trap temperature of 273 K whereas GBAR's cryo-cooled trap can produce positron plasmas as cold as 10 K via cyclotron radiation cooling. The trap temperature in our simulation is achieved using room-temperature nitrogen buffer gas for cooling~\cite{Surko1989}. The externally imposed $\omega_r$ is roughly the same as GBAR's at around 3 MHz. The only constraint on $\omega_r$ is that it is much less than the cyclotron frequency $\Omega_c$. The shape of the positron plasma is well-approximated as a uniform density cylinder with radius $r_p$ and length $l_p$~\cite{Prasad1993}. Appendix~\ref{sec:AppA} provides further details on the distribution of the positrons in the trap.
\begin{table}[!htb] 
\begin{center}
\begin{tabular}{llr}
\hline
Parameter              & Symbol              & Value \\ \hline
Trap radius            & $r_w$               & 4 cm \\
Trap length            & $l_w$               & 10 cm \\
Magnetic field         & $B$                 & 1 T \\
$e^+$ plasma radius    & $r_p$               & 1.3 mm \\
$e^+$ plasma length    & $r_l$               & 5 cm \\
Temperature            & $T$                 & 273 K \\
Number of positrons    & $N$                 & $10^8$ \\
Space charge potential & $\Delta\phi$        & 22.4 V \\
Debye length           & $\lambda_D$         & 60.6 $\mu$m \\
Cyclotron frequency    & $\Omega_c$          & 175.6 GHz \\
Rotation frequency     & $\omega_r$          & 3.2 MHz \\
Transverse emittance   & $\varepsilon_{x,y}$ & 0.11 $\mu$m rad \\
Longitudinal emittance   & $\varepsilon_z$ & 158 cm eV/c \\

\end{tabular}
\caption{Parameters used to define the initial plasma distribution inside the trap. }
\label{tab:parameters}
\end{center}
\end{table}

\begin{figure} [t]
\begin{center}  
\includegraphics[width=0.9\linewidth]{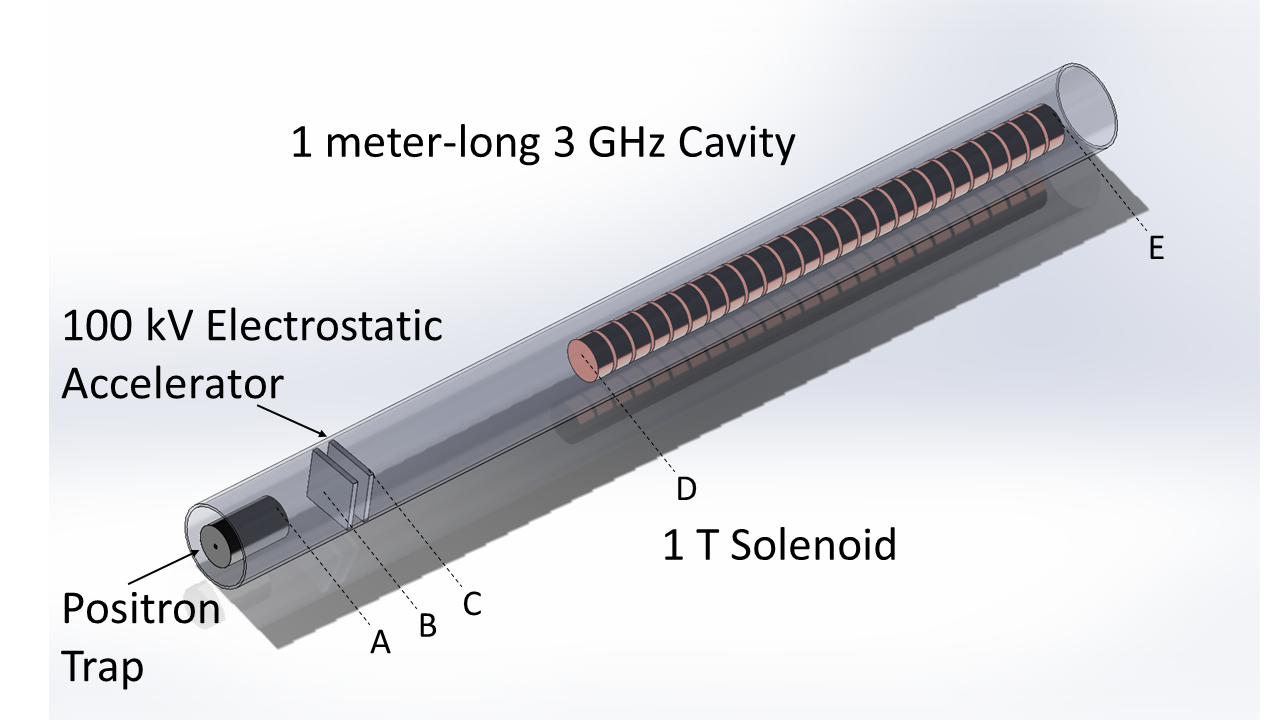}
\caption{\small Depiction of the beamline used in the simulation. The end of the trap is denoted by A, the ends of the electrostatic accelerator are denoted by B and C, and the ends of the 3 GHz linac are denoted by D and E. The entire beamline is contained within a 1 T solenoid.} 
\label{fig:beamline}
\end{center}  
\end{figure}

\section{Analytic Expression for the Trap Emittance}\label{emit}

\begin{figure*}
\begin{center}
    \includegraphics[width=1\linewidth]{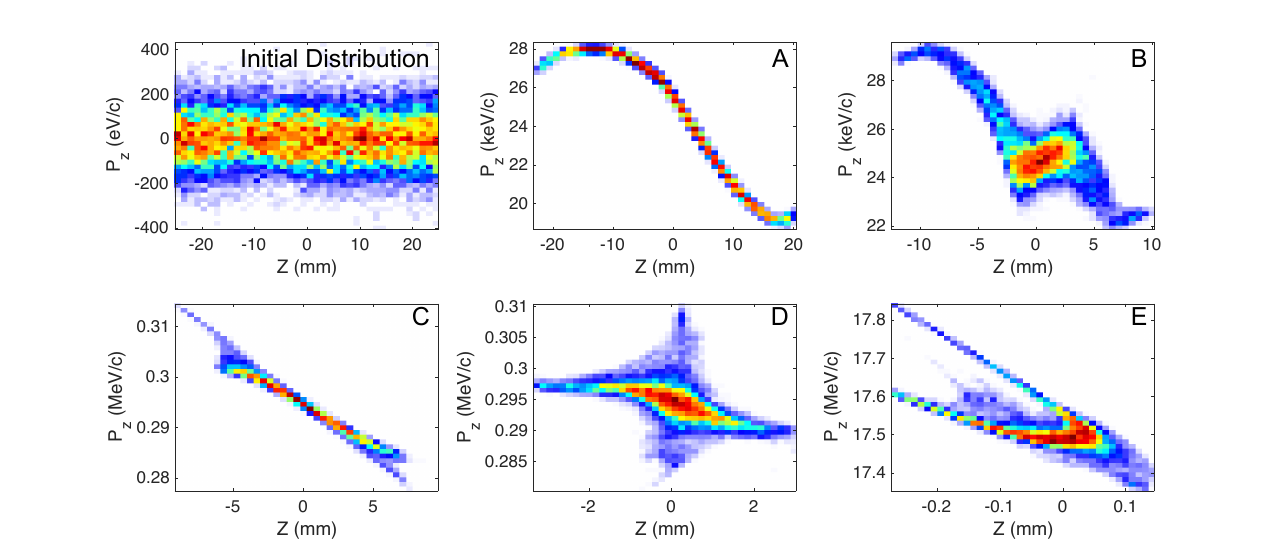}
\caption{\small Longitudinal phase space at demarcated positions along the beamline. The initial distribution corresponds to the beam inside the trap. Positions A through E correspond to the start and end of accelerator components described in Figure~\ref{fig:beamline}.}
\label{fig:phase_spaces}
\end{center}
\end{figure*}

The positrons in the trap form a cold, dense cylinder of charge. Beams that are produced in large magnetic fields, such as those produced by photocathodes in solenoidal fields, have internal angular momentum~\cite{Brinkmann2001,Sun2004}. Upon extraction from the solenoidal field, the beam will acquire an effective emittance proportional to the angular momentum of the beam given by~\cite{PhysRevSTAB.6.104002}
\begin{equation}
    \mathcal{L} = \frac{eB\sigma_r^2}{2mc}.
\end{equation}
In our simulation, we set $B=1$ T and $\sigma_r = 0.65$ mm, and find $\mathcal{L}\approx 124$ $\mu$m rad. 


The relationship between the thermal emittance and effective emittance of the beam is described in Appendix~\ref{sec:AppB}. For our model shown in Fig.~\ref{fig:beamline}, a 1 T solenoid encompasses the entire beamline. The dynamics of the positron bunch in the beamline are unaffected by the beam's angular momentum, which only becomes apparent at the exit of the solenoid.

The thermal emittance of the positron beam in the trap is determined by the plasma temperature and radius of the plasma cylinder. Starting from the standard equation for normalized emittance 
\begin{equation} \label{fullemit}
    \epsilon_n = \frac{1}{m c}\sqrt{\langle x^2\rangle \langle p_x^2 \rangle - \langle x p_x \rangle^2}
\end{equation}
we derive an analytic expression for the transverse emittance in a single plane (here we consider the $x$-plane) of a positron beam at rest in the electrostatic trap. The individual positrons move on cyclotron orbits fixed to field lines and the positron plasma rotates coherently at the rotating wall frequency $\omega_r \ll \Omega_c$. The transverse velocity of the particles is given by the plasma temperature plus a small $r$-dependent correction due to the rotation of the plasma cylinder. The thermal velocity at $T = 273$ K is $v_{th} = 6.4\times10^4$ m/s, whereas the average rotational velocity is $v_\theta = \omega_r \sigma_r = 2.1\times10^3$ m/s. There is no $x-p_x$ correlation and the single-plane transverse emittance reduces to $\epsilon_x = \sigma_x\sigma_{px}/mc$.


With $v_{th}\gg v_\theta$, the momentum spread is 
\begin{equation} \label{pxdef}
    \sigma_{px} = \sqrt{m k_B T},
\end{equation}
and $\sigma_x$ is derived from the uniform positron density extending out to the edge of the plasma cylinder $r_p$
\begin{equation} \label{xcentroid_exp}
    \sigma_x^2 = \frac{r_p^2}{4}.
\end{equation}
Utilizing Equation~\ref{dens_omega} and the finite plasma length $L_p$, we can rewrite $r_p$ purely in terms of trap parameters
\begin{equation} \label{rpdef}
    r_p = \sqrt{\frac{q N}{2 \pi \omega_r \epsilon_0 B L_p}},
\end{equation}
which gives
\begin{equation} \label{xtotsimp}
    \sigma_x^2 = \frac{q N}{8 \pi \epsilon_0 B \omega_r L_p}.
\end{equation}
Combining equations \ref{xtotsimp}, \ref{pxdef}, and \ref{fullemit}, we derive an equation for the normalized, thermal beam emittance defined solely in terms of trap parameters and bunch charge
\begin{equation} \label{trap parouremit}
    \epsilon_{th} = \frac{1}{m c}\sqrt{\frac{q N m k_B T}{8 \pi \epsilon_0 B \omega_r L_p}}.
\end{equation}
For the parameters in our simulation, we find a single-plane thermal emittance of 0.11 $\mu$m rad, which is comparable to or better than the performance of state-of-the-art photocathode guns.

Similarly, considerations apply to the longitudinal emittance $\varepsilon_z = \sigma_z\sigma_{pz}$. The bunch length $\sigma_z = r_l/\sqrt{12} = 1.44$ cm is determined by the placement of the trap electrodes and $\sigma_{pz}$ is again determined by the plasma temperature. We find $\varepsilon_z = 158$ cm eV/c.

\section{Beamline Design and Simulation}

Figure~\ref{fig:beamline} illustrates the beamline used to longitudinally compress and accelerate the beam. The entire beamline is encapsulated by a 1 T solenoid. The simulations of the beamline were performed with the General Particle Tracer (GPT) code~\cite{DeLoos1996}. The beam begins in the electrostatic trap with zero longitudinal energy. The initial bunch distribution is a uniform cylinder~\cite{RevModPhys.87.247}, and the longitudinal extent of the beam is defined by the position of the trap electrodes. The beam in the trap has a bunch length $\sigma_z$ = 14.4 mm (50 mm uniform distribution). The bunch length is long compared to millimeter-scale bunches produced by photocathodes, and much longer than the micron-scale bunches required for PWFA experiments. Therefore, the beam must be longitudinally compressed as it is accelerated. Figure~\ref{fig:phase_spaces} shows the evolution of the longitudinal phase space along the beamline.

\begin{figure} [!htb]
\begin{center}  
\includegraphics[width=0.9\linewidth]{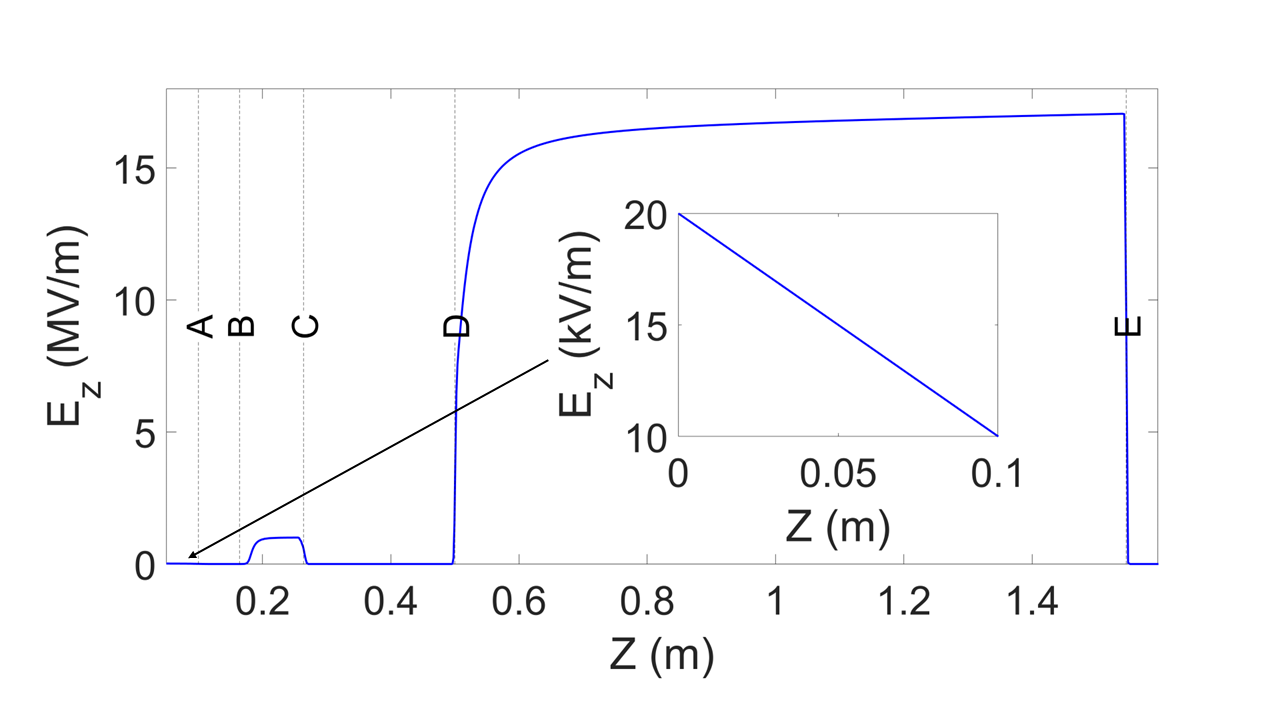}
\includegraphics[width=0.9\linewidth]{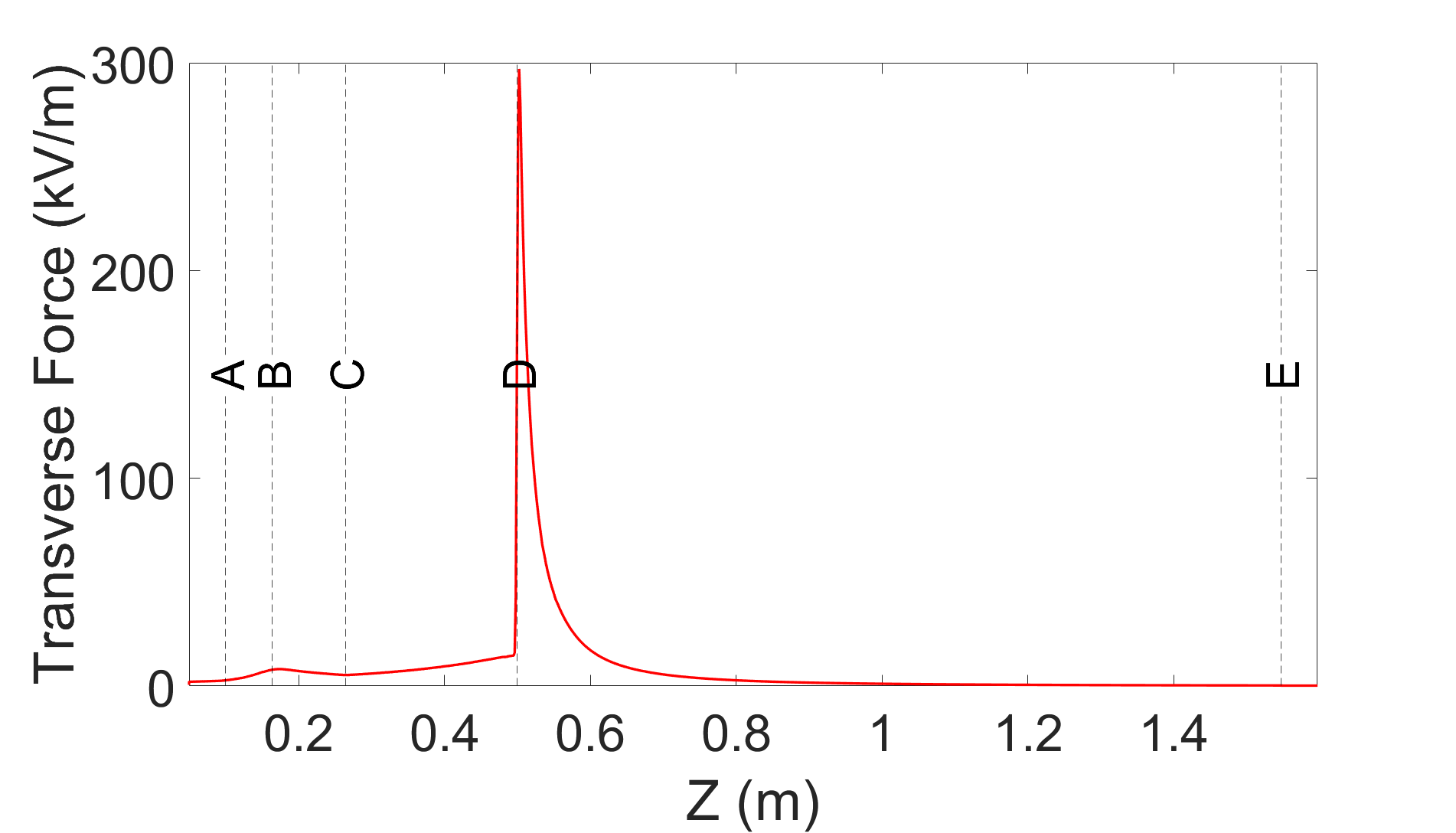}
\caption{Top: Longitudinal field experienced by the positron bunch along the beamline. Bottom: Transverse force $F_\perp = e(E_r - cB_\phi)$ experienced by a particle at a radius of $\sigma_r/2$ from the central axis. The trap extends from $z=0$ cm to $z=10$ cm (Position A), the electrostatic accelerator extends from $z=16.4$ cm (position B) to $z=26.4$ cm (Position C), and the 3 GHz linac extends from $z=50$ cm (Position D) to $z=1.547$ cm (Position E). The magnetic field $B_z = 1$ T is constant along the length of the beamline. The large spike in the transverse field at the entrance to the 3 GHz linac leads to emittance growth, as depicted in Fig.~\ref{fig:bunchlengthandemit}.\label{fig:Ez}}
\end{center}
\end{figure}

Initial compression and acceleration of the long positron bunch is accomplished with a low-field electrostatic buncher inside the trap. A harmonic bunching potential is applied by ring electrodes, such that they provide an accelerating field that decreases linearly along the bunch from the tail to the head~\cite{Mills1980}. The bunching potential is 10 cm long and the bunch initially occupies the central portion of the potential (2.5 cm to 7.5 cm). The voltage drop across the buncher is 2 kV. Figure~\ref{fig:Ez} shows the longitudinal field $E_z$ as a function of position in the accelerator.

The buncher creates a longitudinal focus 7 cm beyond the end of the trap at a longitudinal position of 17 cm in the simulation, immediately after position B in Fig.~\ref{fig:Ez}. A pulsed, 100 kV electrostatic accelerator extends from 16.4 cm to 26.4 cm (positions B to C). The high voltage pulse is provided by a nanosecond pulse generator. The accelerating pulse is timed with the beam such that the field is applied when the beam is between the two accelerating plates. The beam experiences a uniform accelerating field, but positrons at the back of the bunch experience the field for a longer period of time and gain energy relative to particles at the head of the bunch. The beam exits the electrostatic accelerator traveling roughly half the speed of light and undergoes velocity bunching as it travels toward the rf cavity. The second longitudinal focus is at $z = 0.50$ m with $\sigma_z = 1.3$ mm (position D). At this point, the bunch is short enough for injection into the RF cavity.

\begin{figure} [!htb]
\begin{center}  
\includegraphics[width=0.9\linewidth]{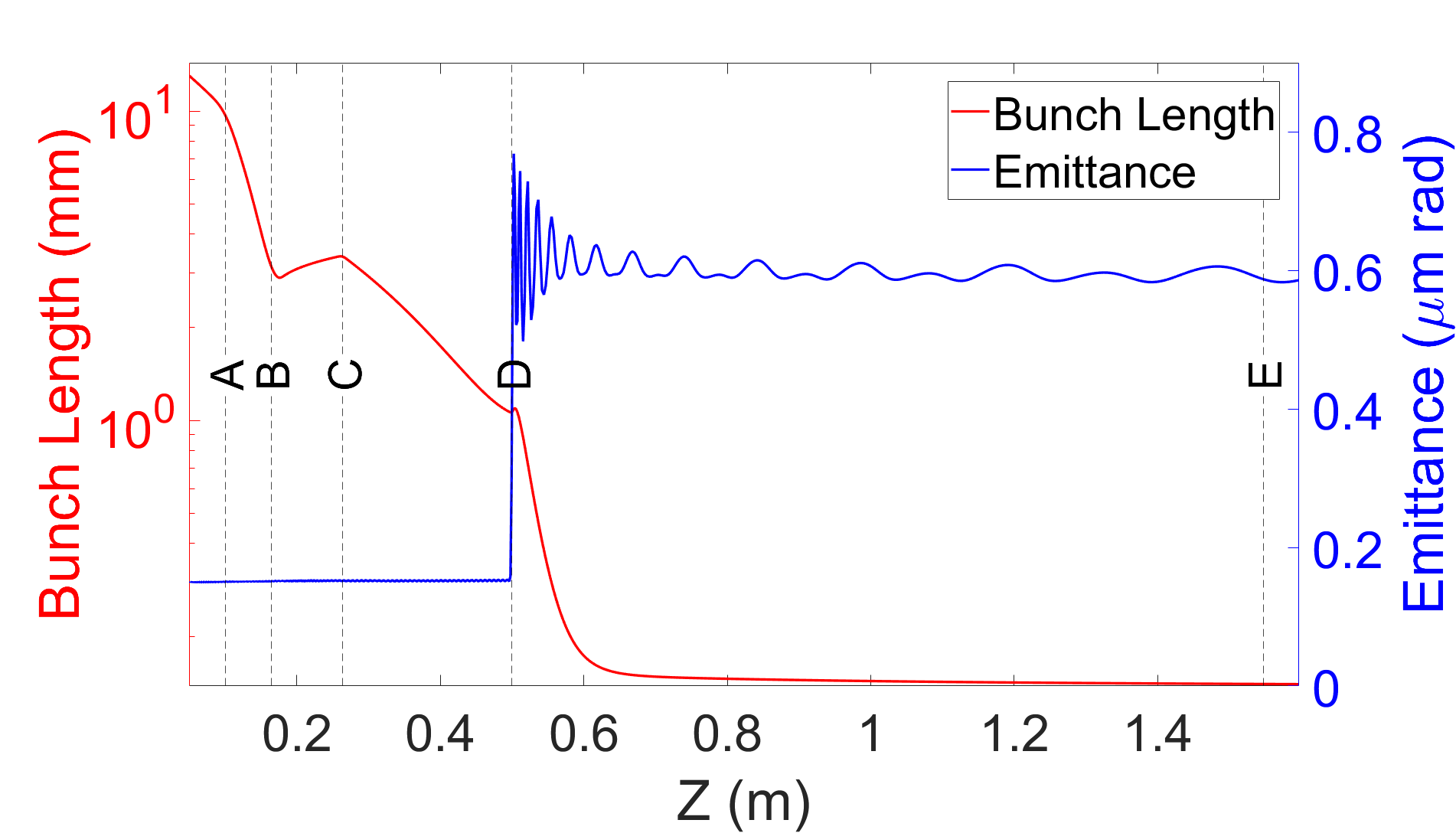}
\caption{\small Bunch length and single-plane thermal emittance (Eq.~\ref{fullemit}) along the beamline. The trap extends from $z=0$ cm to $z=10$ cm (Position A), the electrostatic accelerator extends from $z=16.4$ cm (position B) to $z=26.4$ cm (Position C), and the 3 GHz linac extends from $z=50$ cm (Position D) to $z=1.547$ cm (Position E). The jump in the emittance occurs at the start of the 3 GHz linac.
\label{fig:bunchlengthandemit}}  
\end{center}  
\end{figure}

The entrance to the s-band accelerator structure is located at $z = 0.50$ (position D). The capture phase of the s-band structure is set to both accelerate and longitudinally compress the beam to the final bunch length $\sigma_z$ = 190 $\mu$m and energy of 17.6 MeV. Figure~\ref{fig:bunchlengthandemit} shows the bunch length and emittance along the accelerator. There is an abrupt increase in the emittance from 0.11 $\mu$m rad to 0.60 $\mu$m rad at the start of the s-band cavity due to defocusing rf fields. Further studies will examine the possibility of tailoring the solenoidal magnet field to suppress emittance growth at this location. Table~\ref{tab:slaccomp} shows the output beam parameters. These parameters are comparable to those achieved by the AWAKE electron accelerator~\cite{Kim2020} for injection in a proton beam-driven plasma wakefield. In particular, the bunch length is suitable for injection into plasmas with densities in the $n=10^{14}-10^{15}$ cm$^{-3}$ regime.

\begin{table}[!htb]
\centering
\begin{tabular}{lr}
\hline
Beam parameter               & Value                \\ \hline
Beam energy                  & 17.6 MeV             \\
Beam charge                  & 15.43 pC              \\
Bunch length (rms)           & 190  $\mu$m              \\
Energy spread (rms)          & 0.76\%               \\
Transverse emittance         & 0.60 $\mu$m rad     \\    

\end{tabular}
\caption{Beam parameters at the end of the simulation.}
\label{tab:slaccomp}
\end{table}

\section{Conclusions and Future Work}
The electrostatic trap and beamline described here is capable of producing useful positron beams in a compact footprint. Such a device will enable access to positron beams for accelerator physics studies at universities and national laboratories that currently lack infrastructure for positron beam generation. Although the repetition rate of this positron source is too low for High Energy Physics applications, it is sufficient for studies at PWFA facilities, including the AWAKE facility which produces an experimental shot once every thirty seconds~\cite{Gschwendtner2022}.

Further studies will be undertaken to explore tailored solenoidal magnetic fields that suppress emittance growth at the start of the rf cavity. We also plan to study remoderation of the positron beam to remove intrinsic angular momentum at the cost of reduced bunch charge~\cite{Mills1980b}. The brighter positron beams produced by remoderation may prove useful as a complement to Ultrafast Electron Diffraction (UED) experiments~\cite{Filippetto2022} where the positive beam charge can be used to reduce systematics when used in tandem with electron beams. UED experiments and PWFA studies with positrons require sub-picosecond bunch lengths, which are delivered by our proposed beamline. The ultimate application of this technology would be a positron source for a damping ring-free collider~\cite{Xu2023}. This would require multiplexing of the compact positron source. Multiplexing of positron sources has been previously considered to meet the demands of the NLC collider concept~\cite{Tang}. However, given the repetition rate of existing compact positron sources, this would require thousands of sources operating simultaneously, so research in this direction should focus on increasing the repetition rate of a single source. 

\section{Acknowledgements}
Many individuals helped to provide background on positron sources for this project. We thank Dirk Peter Van Der Werf, Samuel Niang, and Laszlo Liszkay for showing us the GBAR experiment at CERN.  Thank you to David Cooke, David Cassidy, Allen Mills, and Cliff Surko for background on positrons from electrostatic traps. Thank you to Pietro Musumeci for background on UED systems. Klaus Floettman and Bas van der Geer provided input on simulations in ASTRA and GPT, respectively. Thank you to the AWAKE electron source group Seongyeol Kim, Mohsen Dayyani Kelisani, Steffen Doebert, and Edda Gschwendtner from CERN for their useful discussions and support. Work supported by the U.S. Department of Energy, United States under Contract No. DE-AC02-76SF00515.

\appendix

\section{Beam Distribution inside the Penning-Malmberg Trap}\label{sec:AppA}
The positron beam distribution is defined by the parameters listed in Table~\ref{tab:parameters}. Once the external trap parameters are defined (e.g. $B$-field, trap length, rotation frequency), the only free parameter is the number of positrons in the trap $N$. The source of the positron beam, whether that be a radioisotope or a linac-based source, does not affect the distribution of positrons in the trap. This is because the positrons must pass through a solid neon moderator which reduces the kinetic energy of the positrons below the trap voltage threshold~\cite{Mills1980}. The trap voltage threshold is set by Equation~\ref{eq:depth}. The only source-dependent effects on the positron beam parameters are the total positron rate ($e^+/s$) and the polarization of the positrons.

A multistage buffer gas trap is used to reduce the temperature of the positrons in the trap through interactions with room-temperature nitrogen gas~\cite{Surko1989}. The phase space distribution of the finite temperature plasma at equilibrium inside the trap is given by~\cite{Danielson2015}
\begin{equation}
    f(r,z,\vec{v}) = \frac{n(r,z)}{(2\pi k_B T/m)^{3/2}}\exp{\left[-\frac{m(\vec{v}+\omega_r r\hat{\theta})^2}{2k_B T}\right]},
\end{equation}
with $\omega_r$ the rotation frequency of the plasma. The density of the plasma is given by
\begin{equation}
    n(r,z) = n_0 \exp{\left[-\frac{q\phi_{eff}(r,z)}{k_B T}\right]},
\end{equation}
with $\phi_{eff}$ given by
\begin{equation}\label{eq:phi_eff}
    q\phi_{eff} = \frac{1}{2}m\omega_r(\Omega_c - \omega_r)r^2+q\phi_{sc}(r,z)+q\phi_{ext}(r,z)
\end{equation}
for cyclotron frequency $\Omega_c$. $\phi_{ext}(r,z)$ is the potential of the trap electrodes and $\phi_{sc}(r,z)$ is the space-charge potential of the single-species plasma. When the trap is empty, the axial electrodes establish a longitudinal trapping field, but as a consequence of Laplace's equation $\nabla^2\phi_{ext} = 0$, the field is saddle-shaped, i.e. defocusing in $r$ at the center. Further, the space charge force of the single-species plasma is also repulsive in $r$. However, the $\hat{r}$-directed self-field of the plasma induces an $\vec{E}\times\vec{B}$ drift rotation in the solenoidal field of the trap. Radial confinement occurs because of $\vec{v}\times\vec{B}$ Lorentz force of the plasma in the trap, which counters the repulsive space charge force. The $r^2$ dependence of the first term in Eq.~\ref{eq:phi_eff} is the same as if the rotation through a magnetic field was replaced with a uniform cylinder of negative charge with density $n_0$~\cite{Malmberg1977}. The positrons in the trap assume a uniform cylindrical distribution with density decaying exponentially to zero over a Debye length~\cite{Davidson1970,Prasad1979}.

While the positron beam is well-approximated by a uniform cylinder distribution in an idealized scenario, non-uniformities may emerge due to instabilities in the plasma. Diocotron waves with azimuthal mode number $m = 1$ are the most common source of instabilities, as higher mode numbers are strongly damped~\cite{white_reswall}. The diocotron mode has negative energy, so energy dissipation can lead to growth. Different methods for controlling unwanted diocotron mode growth have been developed, including feedback damping techniques~\cite{Hollman_confinement}.

\section{Angular Momentum and Effective Emittance}\label{sec:AppB}
The positron beam is cooled in a strong magnetic field and individual positrons undergo cyclotron orbits along the solenoidal field lines. If the beam is extracted non-adiabatically from the solenoidal field, the positrons will receive an azimuthal kick and their velocity $v_\theta$ will increase. The magnitude of this effect is described by the angular momentum of the beam $\mathcal{L}$. In this appendix, we provide a description of $\mathcal{L}$ and it's relationship to the thermal emittance $\varepsilon_{th}$.

Following the formalism in Ref~\cite{PhysRevSTAB.6.104002}, we define the transverse beam $\mathbf{\Sigma}$ matrix as
\begin{equation}
    \mathbf{\Sigma} = 
        \begin{bmatrix}
        \langle X\tilde{X} \rangle & \langle X\tilde{Y} \rangle \\
        \langle Y\tilde{X} \rangle & \langle Y\tilde{Y} \rangle
        \end{bmatrix},
\end{equation}
with 
\begin{equation}
    \langle X \tilde{X} \rangle = 
        \begin{bmatrix}
        \langle x^2 \rangle & \langle xp_x \rangle \\
        \langle xp_x \rangle & \langle p_x^2 \rangle
        \end{bmatrix},
\end{equation}
and
\begin{equation}
    \langle X \tilde{Y} \rangle = 
        \begin{bmatrix}
        \langle xy \rangle & \langle xp_y \rangle \\
        \langle yp_x \rangle & \langle p_xp_y \rangle
        \end{bmatrix}.
\end{equation}
The transverse emittance $\varepsilon_{4D}$ describes all four dimensions of the transverse phase space and is given by
\begin{equation} \label{4Demit}
    \varepsilon_{4D} = \det(\mathbf{\Sigma}) = \varepsilon_{eff}^2 - \mathcal{L}^2,
\end{equation} 
where $\varepsilon_{eff}$ is the effective emittance in one plane and angular momentum $\mathcal{L} = \frac{1}{2mc}\langle{xp_y-yp_x\rangle}$. 

The thermal emittance is related to the full transverse emittance by $\varepsilon_{th} = \sqrt{\varepsilon_{4D}}$, and the effective single-plane emittance is
\begin{equation}
    \varepsilon_{eff} = \sqrt{\varepsilon_{th}^2+\mathcal{L}^2}.
\end{equation}
The effective single-plane emittance will be dominated by angular momentum when $\mathcal{L}\gg \varepsilon_{th}$. Intuitively, this means that although the volume of the beam in phase space $\varepsilon_{th}$ is small, there are no projections of the beam phase space into the $x-y$ plane such that $\varepsilon_{x}= \varepsilon_{th}$ and $\varepsilon_{y}= \varepsilon_{th}$. However, it is possible to manipulate the beam to minimize either $\varepsilon_{x}$ or $\varepsilon_{y}$ and produce a flat beam~\cite{PhysRevSTAB.6.104002}. The flat-beam transformation can be most easily understood using the concept of eigenemittances described in~\cite{PhysRevSTAB.14.050706,PhysRevAccelBeams.25.044001}. Such beams may be useful for tests of Linear Collider transport systems which employ flat beams from damping rings.


\section{GPT Simulations}


The 3 GHz linac and solenoidal fields are specified using the standard GPT elements ``trwlinbm" and ``Bzsolenoid", respectively. These element include fringe field effects. We created custom elements for the electrostatic trap and bunch compressor using ``map1D\_E''. Only the on-axis fields were specified and the remaining components are calculated from a second-order Taylor expansion of the fields near the beam axis.

\begin{table}[!htb]
\centering
\begin{tabular}{lr}
\hline
GPT simulation parameter     & Value                \\ \hline
Macroparticles               & 20000        \\
Accuracy parameter $\alpha$          & 9          \\
Grid resolution              & 92 $\mu$m             \\

\end{tabular}
\caption{Beam parameters at the end of the simulation.}
\label{tab:GPTparams}
\end{table}

Convergence was demonstrated by adjusting the number of macro particles, grid resolution, and the accuracy parameter $\alpha$. Table~\ref{tab:GPTparams} shows the configuration used for the converged GPT simulations. The accuracy parameter alters the size of the timestep to ensure that results are accurate to one part in $10^\alpha$. Our simulations were particularly sensitive to $\alpha$. Small values of $\alpha$ lead to unphysical leakage of the particles from the trap and an accompanying emittance growth. With $\alpha = 9$, this effect was suppressed.

\bibliographystyle{apsrev}
\bibliography{main}
\end{document}